\documentclass[useAMS,usenatbib]{mn2e}

\usepackage{graphicx} 
\usepackage{comment}

\newcommand{\oiii}{[O\,{\sc iii}]}
\usepackage{amssymb}

\newcommand{\nai}{Na\,{\sc i}}

\begin{document}
\title[On Disentangling IMF Degeneracies in Integrated Light]{On
  Disentangling Initial Mass Function Degeneracies in 
  Integrated Light}
\author[Baitian Tang and Guy Worthey]{Baitian Tang$^{1}$\thanks{E-mail:
baitian.tang@email.wsu.edu (BT)} 
and Guy Worthey$^{1}$\thanks{E-mail:
gworthey@wsu.edu (GW)}\\
$^{1}$Department of Physics and Astronomy, Washington State
University, Pullman, WA 99163-2814, USA\\}

\date{Accepted  . Received  ; in original form 2014}

\pagerange{\pageref{firstpage}--\pageref{lastpage}} \pubyear{ }

\maketitle

\label{firstpage}

\begin{abstract}

The study of extragalactic integrated light can yield partial
information on stellar population ages, abundances, and the initial
mass function (IMF). The power-law slope of the IMF has been studied
in recent investigations with gravity-sensitive spectral indicators
that hopefully measure the ratio between KM dwarfs and giants. We
explore two additional effects that might mimic the effects of the IMF
slope in integrated light, the low mass cutoff (LMCO) and a variable
contribution of light from the asymptotic giant branch (AGB). We show
that the spectral effects of these three (IMF slope, LMCO, AGB
strength) are very similar and very subtle compared to age-abundance
effects. We illustrate parameter degeneracies and covariances and
conclude that the three effects can be disentangled, but only in the
regime of very accurate observations, and most effectively when
photometry is combined with spectroscopy.

\end{abstract}


\section{Introduction}
\label{sect:intro}

Deriving astrophysical parameters from integrated-light observables is
a widely accepted practice in the field of Galactic and extragalactic
research. However, increasing evidence shows that single-burst,
single-composition stellar populations oversimplify the underlying
stellar systems \citep{Gratton2012, Kaviraj2007}. Additional
parameters in the modelling process, such as: multiple burst-age
stellar populations \citep{Trager2000b, Trager2005, Goudfrooij2011},
metallicity and helium abundance variation \citep{Norris2004,
  LeeYW2005}, initial mass function (IMF) variation
\citep{Weidner2013a, Bekki2013, Chabrier2014}, and chemical abundance
variation \citep{Dotter2007,Lee2009} were investigated in hopes of
reconciling various contradictions between observation and
theory. This group explored metallicity-compositeness by varying the
abundance distribution function (ADF; d$M$/d[M/H]; the mass fraction
of the stellar population at each [M/H]) effect in
\citeauthor{Tang2014} (2014, Paper I). The ADF shape is a gentle rise
at low abundance, a peak, and a steeper fall-off at high abundance. By varying the widths of the ADFs we discovered ``red lean'' and ``red
spread'' phenomena. ``Red lean'' means that a narrower ADF appears
more metal-rich than a wide one, and ``red spread'' describes that the
spectral difference between wide and narrow ADFs increases as the ADF
peak is moved to more metal-rich values.

We continue to explore in this paper additional underexplored effects,
this time ones that might endanger measurements of the IMF slope in
integrated stellar population (SP) models. The IMF is of importance in
stellar and extragalactic astronomy, affecting galaxy luminosity
evolution, star formation, chemical evolution, and mass budget between
normal and dark matter. The IMF regulates the mass distribution of the
stellar population, and thus impacts the luminosity function, mass to
light (M/L) ratio, and number of stellar remnants.  However, direct
IMF slope derivations that counts individual stars are limited by
observational capabilities, uncertainties concerning the
mass-luminosity relation, stellar evolution, dynamical evolution,
binary fraction, and many other factors.  Recently reemerged
gravity-sensitive spectral lines (e.g., CaT, \nai, and Wing-Ford band)
show bright future applications for indirectly estimating the IMF
slopes of unresolved galaxies \citep{Cenarro2003,
  VanDokkum2010,Conroy2012b}.  Gravity-sensitive spectral lines
measure the dwarf/giant ratio. Our stance is that this ratio could be
altered by effects other than the IMF slope. 

First, a low-mass cut-off for the IMF
(LMCO)\footnote{Throughout this paper, IMF LMCO is called LMCO for
  short.}, exists in every set of SP models. It is the mass limit of
stars included at the lower mass end --- any star with mass smaller
than this limit is assumed to have no contribution in the SP
models. Often, the LMCO is a pragmatic choice dictated by the choice
of stellar evolutionary isochrones that go into the SP models. It is
readily appreciated that the LMCO is closely related to the fraction
of low mass stars, and thus the dwarf/giant ratio.  Therefore, we may
encounter degeneracy when determining IMF 
slope and LMCO simultaneously. The LMCO is variable among different SP
models found in the literature. For example, the Padova isochrones
\citep{Bertelli2008,Bertelli2009} set 0.15 $M_{\odot}$ as the LMCO,
while the composite isochrones of \citet{Conroy2012a} sets 0.08
$M_{\odot}$ as the LMCO. As a result, the derived IMF slope values may
change when swapping between SP models.

Second, we note the possibility that the dwarf/giant ratio may come as
easily from modulating the number of giants as modulating the number
of dwarfs. The bolometrically brightest giants are asymptotic giant
branch (AGB) stars. These evolved stars have long been known to be
difficult to constrain in most SP models
\citep{Conroy2010,Girardi2010,Girardi2013} due to uncertain stellar
evolution (in turn due to uncertain rules about mass loss in giants)
and also difficult to empirically constrain due to counting statistics
\citep{Frogel1990,Santos1997,Bruzual2003,Salaris2014}.  We suggest a
plausible, hypothetical trend, \textit{ADF-AGB-IMF masquerading}, in
which increased metallicity causes cooler giants, which causes
increased mass-loss, which causes fewer AGB stars, which resembles an
increase in IMF slope.

In this paper, we study spectral and photometric variations in
response to simultaneous changes of IMF slope, LMCO, and AGB strength
($\S$\ref{sect:main}).  We show that the degeneracies can be
marginally lifted for old, metal-rich populations, but the
degeneracies are still firm in young populations ($\S$\ref{sect:old}
\& $\S$\ref{sect:young}). Next, we dig into the mechanisms behind all
these variations by studying the numbers of stars, luminosities, colors,
and spectral indices of each evolutionary phase ($\S$\ref{sect:reason}).

Bottom-heavy IMFs are indicated in local massive and metal-rich
elliptical galaxies \citep{Cenarro2003, VanDokkum2010, Conroy2012b,
  Cappellari2012}. On the other hand, star-formation theories such as
\citet{Larson1998,Larson2005} and \citet{Marks2012} imply the
metal-free early Universe favors a top-heavy IMF.
\citet{Weidner2013a} points out a time-independent bottom-heavy IMF
generates too little metal and fewer stellar remnants than observed.
The hypothesis that IMF steepens as the Universe evolves 
will change the light ratio between
metal-poor and metal-rich stars in a similar way as the ADF effect, thus
we call here \textit{ADF-IMF coupling}. 
In $\S$\ref{sect:coupling}, we build rough models with ADF-IMF
coupling to explore the ramifications, at least qualitatively. The
coupled models appear more metal-rich than the noncoupled models, due
to the suppression of metal-poor stars, resembling the \textit{red
  lean} effect.

Finally, we test parameter recovery using a Monte Carlo approach. 
The mean recovered values agree with the input values inside the
error range without alarming systematics ($\S$\ref{sect:recover}).  We
uncover covariances among the IMF slope, LMCO, and AGB effects when
combined with the more usual age and metallicity effects.  Though the
magnitudes of the IMF-related effects are smaller than the latter
effects, these two groups of effects vector almost orthogonally
($\S$\ref{sect:all}), and we prognosticate bright hopes for
disentangling all of these stellar population parameters.  A summary
of our results are given in $\S$\ref{sect:con}.

\section{IMF slope, LMCO, and AGB}
\label{sect:main}

\subsection{Model description}
\label{sect:mo}

A new version of old integrated-light models
\citep{Worthey1994a,Trager1998} is adopted.  The new models use a new
grid of synthetic spectra in the optical \citep{Lee2009} in order to
address the effects of changing the detailed elemental composition on
an integrated spectrum.  The models retain single burst age and
metallicity as parameters, but were expanded to also include
metallicity-composite populations, and the three IMF-related
parameters we discuss (IMF slope, LMCO, and AGB modulation).

For this work, we adopt the isochrones of
\citet{Bertelli2008,Bertelli2009} using the thermally-pulsing
asymptotic giant branch (TP-AGB) treatment described in
\citet{Marigo2008}. Improving upon \citet{Poole2010}, stellar index
fitting functions were generated from indices measured from the
stellar spectral libraries of \citet{Valdes2004} as re-fluxed by
I. Chilingarian (2014, private communication), \citet{Worthey2014x},
MILES \citep{FalconBarroso2011}, and IRTF
\citep{Rayner2009}, all transformed to a
common 200 km s$^{-1}$ spectral resolution via smoothing if the
spectral resolution was greater or by linear transformation after
index measurement if the resolution was lesser. Index errors were
assigned by iteratively comparing the libraries and seeking error
models of the form $\sigma^2 = \sigma^2_{\rm{floor}} +
C\sigma^2_{\rm{photon}}$, where $\sigma_{\rm{floor}}$ is an error
floor mostly applicable to indices of large wavelength span, and $C$
is a fitting constant. Multivariate polynomial fitting was done in
five overlapping temperature swaths as a function of $\theta_{eff}$ =
5040/T$_{eff}$, log g, and [Fe/H]. The fits were combined into a
lookup table for final use. As in \citet{Worthey1994a}, an index was
looked up for each bin in the isochrone and decomposed into ``index''
and ``continuum'' fluxes, which added, then re-formed into an index
representing the final, integrated value after the summation.

The IMF slope, LMCO, and AGB effects are calculated at two
representative values. The IMF slope\footnote{A power-law function,
  where 2.35 is the \citet{Salpeter1955} slope and 1.70 is a
  bottom-light hypothesis.} is calculated at 2.35 and 1.70, the LMCO
at 0.15 and 0.30 $M_{\odot}$, and the AGB contribution with full and
80\% strength. For modeling, we define the AGB phase as beginning 0.5
mag brighter than the red clump, and continuing to the end of the
star's life.  The number of AGB stars of each mass bin is reduced to
80\% to simulate a weaker AGB strength model. The SP model with IMF
slope of 2.35, LMCO of 0.15 $M_{\odot}$, and full AGB strength is
chosen as the standard model.  Parameter variations from these nominal
values are carried with linear interpolation. Naturally, if the
parameters drift too far from the modeled values they can no longer be
considered realistic.

To illustrate the different effects we vary one parameter at a
time.  That is, the results we display graphically are
partial derivatives ($\partial_{index} /\partial_{effect} $).

\begin{figure*}
\includegraphics [width=0.9\textwidth]{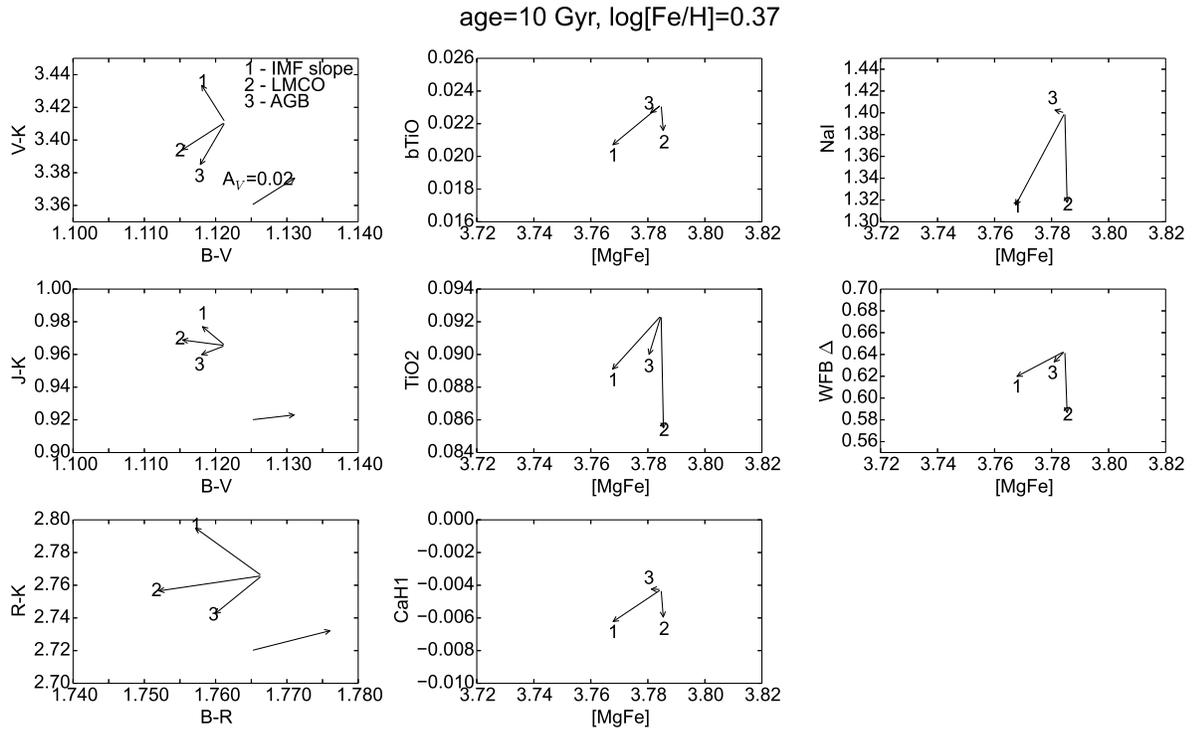} 
\caption[Color-color and index-index plots for old, metal-rich
population]
{Color-color and index-index plots for an old, metal-rich population. The
IMF slope, LMCO, AGB effects are labelled as vectors 1, 2, 3,
respectively. At the bottom right of each plot involving photometric colors, an
extinction vector of A$_V=0.02$ mag is sketched.}\label{fig:qs}
\end{figure*}

\subsection{An Old, Metal-rich Population}
\label{sect:old}

Since most of the elliptical galaxies in the local Universe are old
and metal-rich, we pick age $=10$ Gy, and $\log(Z)=0.37$ to mimic a
typical elliptical galaxy.  In the left column of Figure \ref{fig:qs},
we show optical-near infrared (NIR) color-color plots. Note that we
set the same y axis dynamic range (0.1 mag) for all the color-color
plots\footnote{0.01 mag and 0.15 \AA~for magnitude and angstrom unit
  indices, respectively.}. An extinction vector of A$_V=0.02$ mag is
sketched on the bottom right of each plots. We note that (1) the IMF
slope, LMCO, and AGB effects are not parallel vectors, which means it is
practical to lift the degeneracy of a single index with a mix of
different colors.  However, we note that these effects are detectable
at the level of $\sim$0.02 mag. Such accuracies are comparable to
contemporary detection limits and therefore technically challenging by
today's standards. In addition, (2) the ($J-K$) vs. ($B-V$) plot shows
the weakest drifts among the three plots, due to the
wavelength-proximity of these two pairs of filters. This
wavelength-proximity is also the reason for small extinction
vector. We also find that (3) decreasing the IMF slope and AGB
strength tend to change the ($X-K$) colors in the opposite direction.
Interestingly, increasing the LMCO leads to a decrease in ($V-K$) and
($R-K$), but an increase in ($J-K$), implying blue ($J-K$) colors for
stars between 0.15 and 0.30 $M_{\odot}$ ($\S$\ref{sect:reason}). (4)
Finally, using colors alone, dust extinction is almost
indistinguishable from LMCO effects.

In the middle and right columns of Figure \ref{fig:qs}, we plot
[MgFe]\footnote{[MgFe]=$\sqrt{{\rm Mgb}*({\rm Fe}5370+{\rm
      Fe}5335)/2)}$} versus five IMF-sensitive indices.  We introduce
to the Worthey models the optical IMF-sensitive indices bTiO and CaH1
from \citet{Spiniello2014a}, and plan to use them for future
intermediate-redshift IMF research (Tang et al., in preparation). We
notice that (1) the drifts caused by weakening the AGB strength are
the smallest, due to low flux contribution of the AGB stars at the age
of 10 Gyr ($\S$\ref{sect:reason}).
We also see that (2) decreasing the IMF slope tends to
decrease the [MgFe] index (about 0.02 \AA), while increasing the LMCO leads to 
small variation of [MgFe]. In addition, 
(3) the smaller dwarf/giant
ratios induced by the IMF slope and LMCO effects
lead to smaller values for IMF-sensitive indices. 
But the magnitude ratios of these two effects vary for different indices.
The results that the LMCO drifts are larger than the IMF slope drifts in TiO2 and
WFB  $\Delta$ (Wing-Ford band) \citep{Wing1969,Whitford1977,Hardy1988}
suggest caution should be taken when interpreting these 
two indices as a clean IMF slope indicator.

\begin{figure*}
\includegraphics [width=0.9\textwidth]{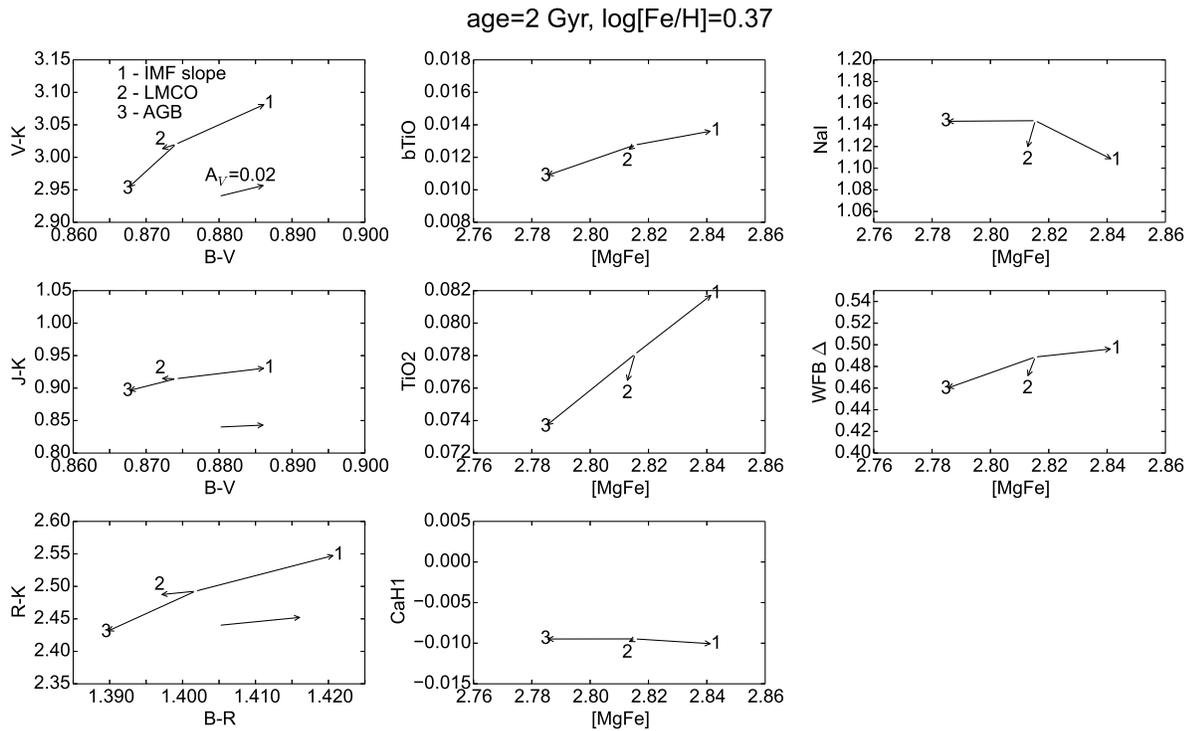} 
\caption[Color-color and index-index plots for young, metal-rich
population]
{Color-color and index-index plots for a young, metal-rich population. The
IMF slope, LMCO, AGB effects are labelled as vectors 1, 2, 3,
respectively. At the bottom right of each color-color plot, an
extinction vector of A$_V=0.02$ mag is sketched.}\label{fig:y}
\end{figure*}

\subsection{A Young, Metal-rich Population}
\label{sect:young}
\citet{Schiavon2006} suggested young elliptical
galaxies with ages of the order of 1 Gyr prevail in the early Universe (z$\sim$0.9).
Studying these galaxies places important constraints on the elliptical
evolution models. Here we pick age
$=2$ Gyr, and $\log(Z)=0.37$ to mimic a young elliptical galaxy.
In the all the plots of Figure \ref{fig:y}, weakening the AGB strength
induces a stronger drift than the old population, due to the high
flux contribution of AGB stars at the age of $0.2-2$ Gyr
\citep{Maraston2006}. To accommodate the large AGB drifts, the 
$y$ axis dynamic ranges are set to 0.25 mag for color-color
plots. Note that this may
causes visual differences between the color-colors plots of Figures
\ref{fig:qs} and \ref{fig:y}. 

In the optical-NIR color-color plots of Figure \ref{fig:y}, (1) increasing
the LMCO leads to negligible signal, while weakening the AGB
strength shows a drift that is  a factor of two larger as the old
population. Another noticeable change 
is (2) the optical colors on $x$ axis. Instead of getting bluer as the old
population does, decreasing the IMF slope of a young population drives
the ($B-V$) and ($B-R$) colors redder, probably due
to stronger post main sequence phases  ($\S$\ref{sect:reason}). 
Comparing with the extinction vectors, (3) the AGB drift and IMF slope drift
should be clearly detectable at 0.02 mag level.

In the middle and right columns, (1) the index increases of bTiO, TiO2,
and WFB $\Delta$ caused by decreasing the IMF slope of a young
population confirm that IMF determination is sensitive to the age
parameter. We also see that (2) the shallower IMF slope model also has greater [MgFe] than
the standard model.  In addition, (3) the AGB and IMF
slope effects vector almost oppositely in most of the Figure
\ref{fig:y} plots.  The partial derivative nature of our results
reveals the degeneracy of IMF slope and AGB strength is still firm for
the young population.  We further discuss the reasons for different drift
directions by inspecting the luminosity weights, colors, and indices of each
evolutionary phase in $\S$\ref{sect:reason}.

\subsection{Colors and Indices Broken into Evolutionary Phases}
\label{sect:reason}

\begin{table*}
\caption{\label{tab1} Colors and Indices of Each Phase for the Old Population}
\begin{center}
\begin{tabular}{lcccc}
\hline \hline
&  &OLD POPULATION& & \\
&PHASE a& PHASE b& PHASE c& PHASE d\\
\hline
(B-V)& 1.46939& 0.69314& 0.93189& 1.48725\\
(V-K)& 4.04228& 1.75775& 2.87041& 4.84277\\ \relax
[MgFe] & 3.77325& 2.34650& 2.83331& 4.98660\\
bTiO& 0.17246&-0.00274& 0.00573& 0.18546\\
TiO2& 0.44303& 0.01672& 0.07097& 0.33609\\
WFB $\Delta$& 2.07944& 0.06787& 0.35408& 1.16696\\

\hline
\end{tabular}
\end{center}
\end{table*}

\begin{figure}

\includegraphics [width=0.5\textwidth]{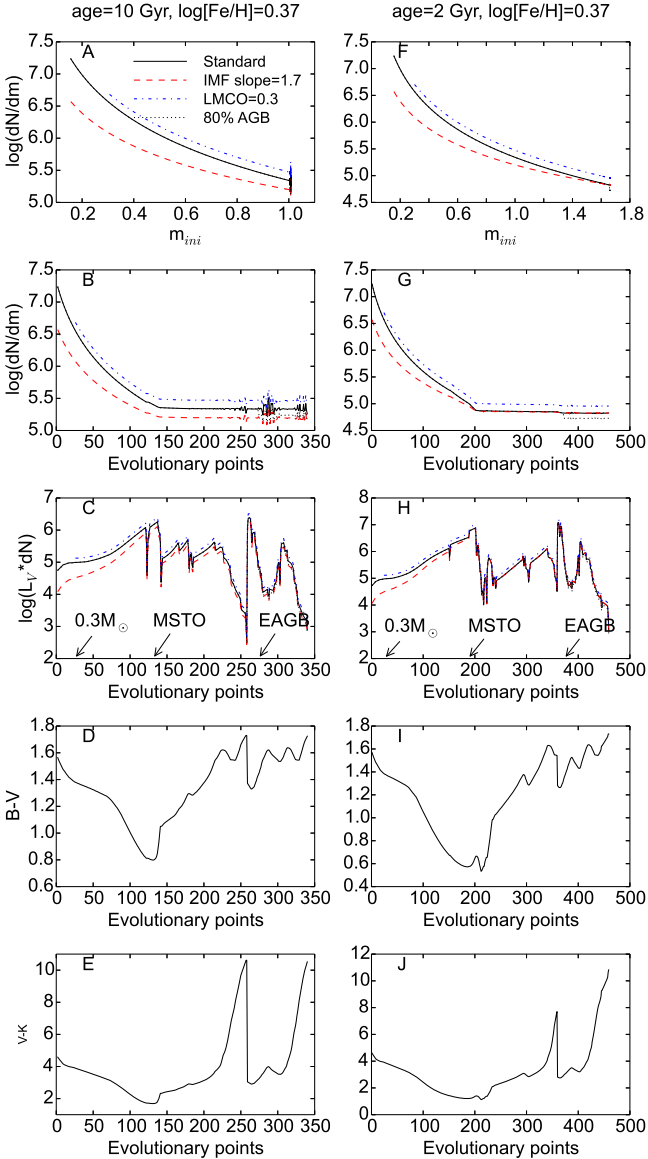} 
\caption[Number of stars, luminosities, and colors vs. evolutionary
points]{The number of stars, luminosities, and colors of the old, and
  young populations are plotted against isochrone evolutionary points or
  m$_{ini}$, where the use of evolutionary points better higlights post
  main-sequence phases.  Four phases are defined as: phase $a$,
$0.15<m_{ini}<0.30~ M_{\odot}$; phase $b$, $m_{ini}=0.30~ M_{\odot}$ to
the main sequence turn-off (MSTO); phase $c$, from MSTO to the beginning
of the AGB phase, or early AGB (EAGB); phase 
d, the AGB phase.}\label{fig:vkm}
\end{figure}

\begin{figure}

\includegraphics [width=0.55\textwidth]{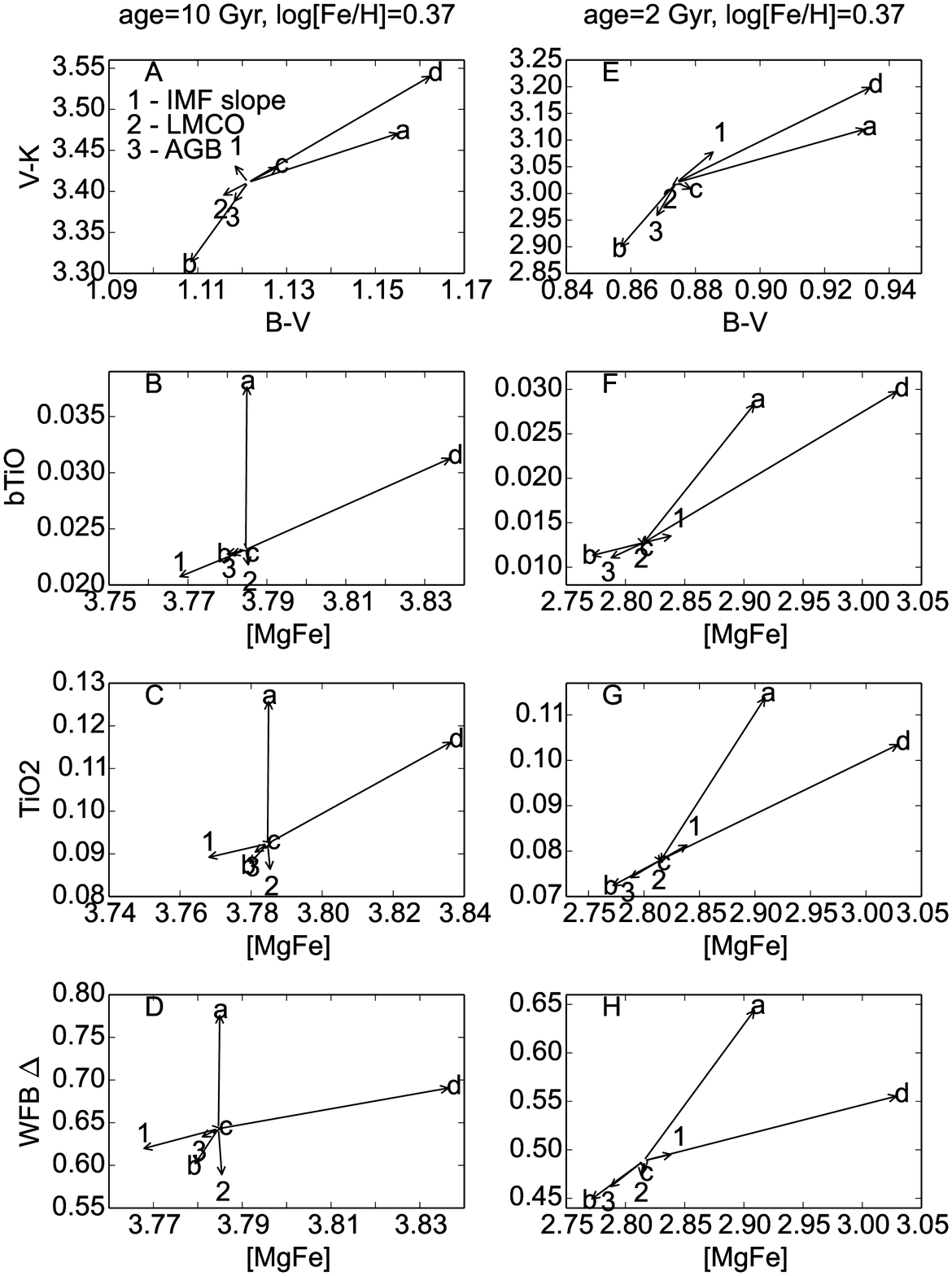} 
\caption[Vectors presenting star number increment in different
phases.]{To illustrate the connections of effects and evolutionary
  phases, we connect the standard models with the contributions of each
  phase. The phase vectors (a,b,c,d) are shrunk to one tenth of the
  original magnitude to accommodate the three effect vectors. The
  vector of each phase is labelled by the corresponding phase
  name.}\label{fig:vec}
\end{figure}

To illustrate the reasons behind the color drifts of three effects, we
show the number of stars, luminosities, and colors in Figure
\ref{fig:vkm}. First, we take a look at the old population. In Figure
\ref{fig:vkm}A, we plot the logarithmic number of stars per initial
mass ($\log(dN/dm)$) versus initial mass ($m_{ini}$). All stars with
$m_{ini} \gtrsim 1.0~M_{\odot}$ have evolved to the stellar remnant
phase, thus they do not contribute to the optical-NIR integrated light
any more.  The shallower IMF model has smaller $dN/dm$ than the
standard model in the range of 0.15 to $\sim 1.0~M_{\odot}$, while the model
with 0.3 $M_{\odot}$ LMCO has more stars in the range of 0.3 to$\sim
1.0~M_{\odot}$, due to a normalization correction (all stellar
population are normalized to the same initial mass, so, for example,
raising the LMCO means that more mass and more light is distributed to
the stars that remain). The 80\% AGB model is
not discernible here, because the AGB stars have very small initial
mass range ($\sim 1.0~M_{\odot}$). To inspect the strong variance of
the post main sequence phases, we use the evolutionary points as $x$
variables instead of $m_{ini}$. The evolutionary points are defined as
points with the same interval arc length along the isochrones in the
HR diagram. Figure \ref{fig:vkm}B clearly shows different number of
stars for the standard model and the 80\% AGB strength model. In
Figure \ref{fig:vkm}C, we find that the post main sequence stars,
though small in number, have major contribution in the $V$ band
integrated light. Then we plot the ($B-V$), and ($V-K$) colors in Figure
\ref{fig:vkm}D and \ref{fig:vkm}E. Obviously, the very low mass stars
and the giants have comparable ($B-V$) colors, but the giants, on
average, show redder ($V-K$) colors. 

\begin{figure*}
\includegraphics [width=1.0\textwidth]{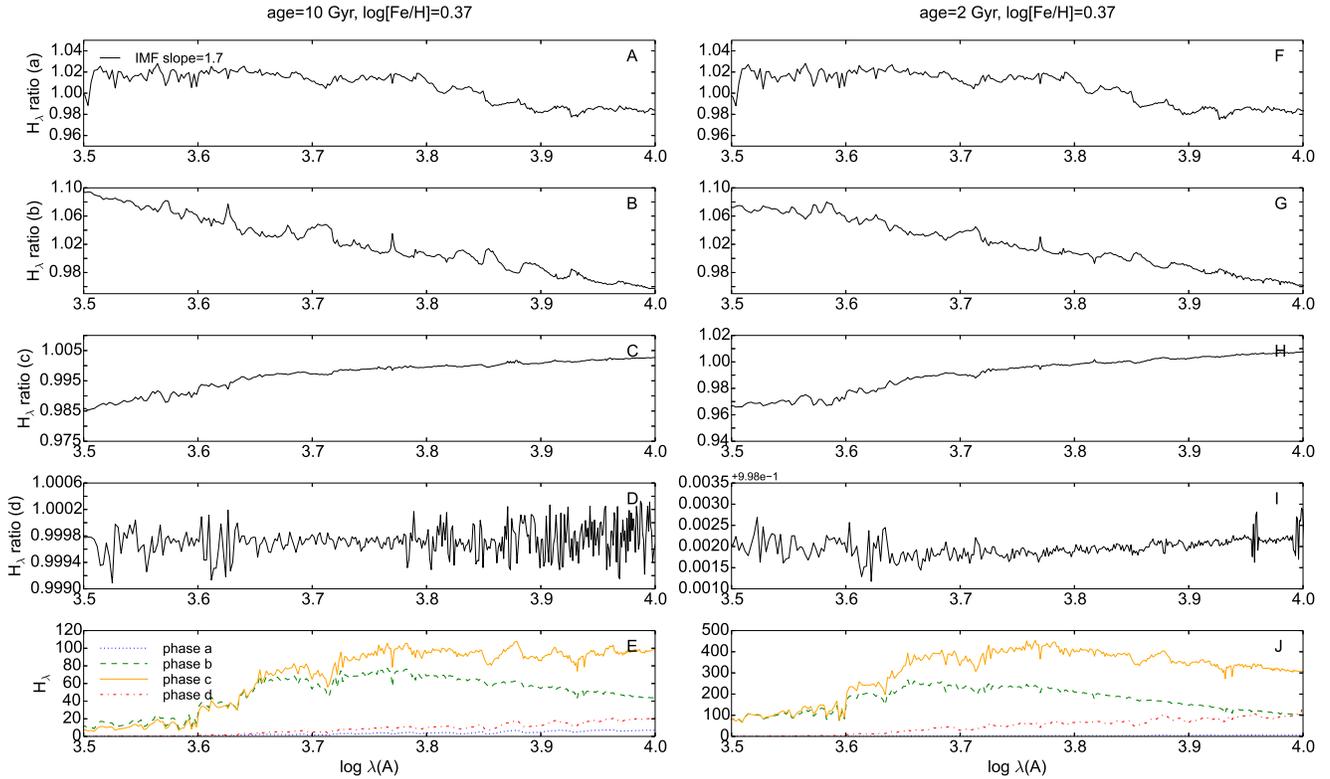} 
\caption[Spectra and Spectral ratios of each phase.]
{The upper four panels show the flux ratio between the shallower IMF
  models and the standard models, for four different phases. The very
  bottom panels plot the spectra of the standard models, 
  subdivided into four phases.}\label{fig:spec}
\end{figure*}

To better quantify the contribution
of different phases, we first define four phases: phase a,
$0.15<m_{ini}<0.30~ M_{\odot}$; phase b, $m_{ini}=0.30~ M_{\odot}$ to
the main sequence turn-off (MSTO); phase c, from MSTO to the beginning
of the AGB phase, or early AGB (EAGB); phase 
d, the AGB phase. Next, we run the models with only one phase on by
nullifying the contribution of all other phases. For example, to study
the integrated light from phase a stars, we set the number of stars in
phase b, c, and d to zero. The colors and indices from
each phase are tabulated in Table \ref{tab1}.  A vector that connects
the standard model and one of these indices (or colors) indicates the drift direction if
more stars in the corresponding phase are added.  In Figure \ref{fig:vec}, these
vectors are shrunk to one tenth of the original magnitude to
accommodate the three drift vectors. In the ($V-K$) vs. ($B-V$) plot, the
phase a vector is in the opposite direction of the LMCO drift, and
similarly the phase d vector has an angle of almost 180 degrees with
the AGB drift. These indicate the LMCO effect
is mainly caused by decreasing the contribution of phase a stars, and
the AGB effect is mainly caused by decreasing the contribution of phase
d stars. These two conclusions might be obvious, but no vector seems
directly related to the IMF slope drift. The IMF slope effect changes
the number of stars of all masses, which means the high-to-low-mass
ratio of each phase is not a constant. Therefore, we need to take the
flux contribution of each phase into consideration. In Figure
\ref{fig:spec}E, we plot the integrated optical spectra of four
phases. The spectra of phase b and c are both luminous, although phase
c stars become brighter than phase b stars as 
wavelength exceeds $\log(\lambda)\approx 3.65$ ($\approx$ 4467~\AA).
In Figure \ref{fig:spec}A, \ref{fig:spec}B, \ref{fig:spec}C, and
\ref{fig:spec}D, we search for the spectral variation signals. The
spectra of each phase are first normalized at $\log(\lambda)= 3.815$. Then the spectrum of the
shallower IMF model is divided by the standard model. The spectral
ratios show shallower IMF model tends to blue the spectrum of phase b
stars, but redden the spectrum of phase c stars.  Therefore, the
bluing and reddening effects compete again each other to determine the
integrated color drifts, depending on the flux ratios of phase b stars
to phase c stars (b/c) at different wavelengths. Let us take the ($B-V$)
vs. ($V-K$) plot as an example. The b/c flux ratio is close to unity at
the $B$ and $V$ bands, but much smaller than 1 at the K band.  Thus the
($B-V$) color is governed almost equally by both the phase b stars and
phase c stars, while the ($V-K$) color is mainly controlled by the phase
c stars that redden the spectrum.  The ($B-V$) color drifts blue due to
the stronger spectral variation of phase b stars (Figure
\ref{fig:spec}B) compared to phase c stars (Figure \ref{fig:spec}C).

\begin{figure*}
\centering
\includegraphics [width=0.8\textwidth]{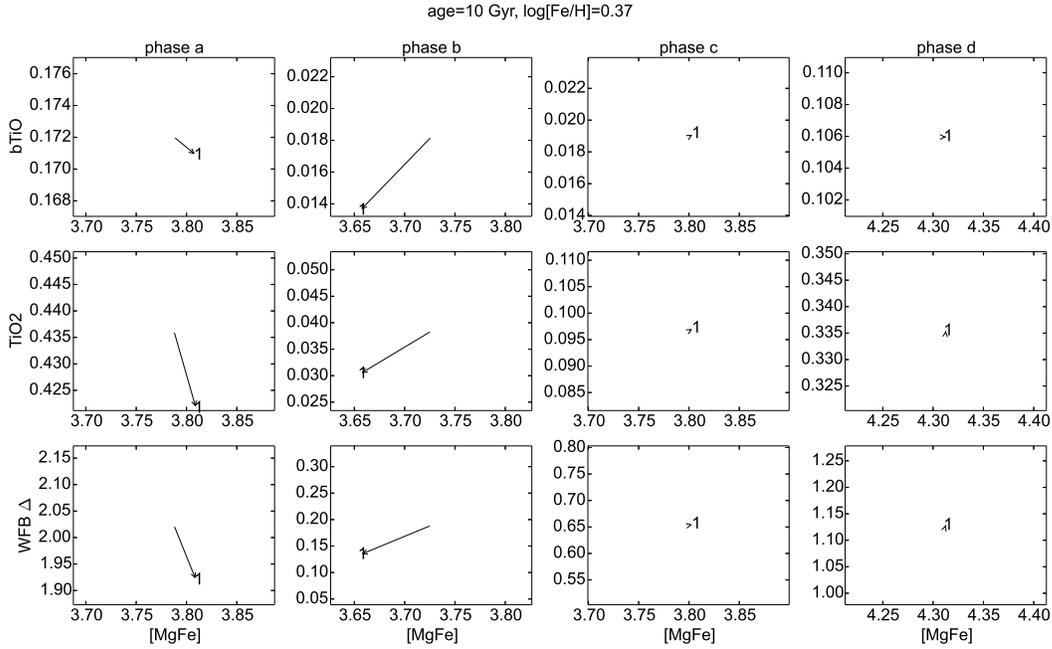} 
\caption[IMF slope drifts of each phase at old age]{IMF slope drifts of each phase at old age.}\label{fig:ind1}
\end{figure*}

In the index-index plots of Figure \ref{fig:vec}B, \ref{fig:vec}C, and \ref{fig:vec}D, similar
conclusions as Figure \ref{fig:vec}A can be drawn for the LMCO and AGB drifts, based on the a
and d vector directions. However, the index variations for the IMF
slope effect cannot be easily determined from Figure
\ref{fig:spec}. To better quantify the response of different phases,
we show the index-index plots of all four phases in Figure \ref{fig:ind1}.
We notice the drift magnitudes of the phase b stars are much greater
than the phase c stars. Since the smallest b/c flux ratio is about
0.5 (at WFB, Figure \ref{fig:spec}E), phase b stars dominate the index variation,
pushing all the indices to lower values.

Next, we switch to the young population. A major difference compared to the
old population is that stars with mass between 1.0 to $\sim$1.66
M$_{\odot}$ still exist (Figure \ref{fig:vkm}F). Two curves of
different slopes intersect at $\sim$1.64 M$_{\odot}$, thus the shallower
IMF model has more AGB stars than the standard model!  Furthermore,
Figure \ref{fig:vkm}H shows longer post main sequence phases than the
old population.  In Figure \ref{fig:vkm}I and \ref{fig:vkm}J, we
clearly see the MSTO hook features near the curve minimum. The hook
feature happens when the star transitions from core burning to shell
burning, with a brief gravitational-energy-dominated moment.  The
($V-K$) color of the RGB tip is smaller than the old
population, due to higher stellar mass and thus higher surface
temperature of the RGB tip stars at young age.  In Figure
\ref{fig:vec}E, \ref{fig:vec}F, \ref{fig:vec}G, and \ref{fig:vec}H,
the directions of vector a and d again testify the previous statement that
the LMCO effect is mainly caused by phase a stars, while the AGB effect
is mainly caused by phase d stars.  To constrain the major contributor
of the IMF slope effect, we plot the spectra and spectral ratios of
all four phases in the right column of Figure \ref{fig:spec}. The
spectrum of phase a is dwarfed by the flux increase of other three
phases. Phase c is the most significant of all, starting to overcome
phase b at $\log(\lambda) \approx 3.53$.  As a result, phase c stars
contribute most of the light in our optical, NIR colors and
indices. Obviously, the spectral bluing and reddening competition of
phase b and phase c stars (Figure \ref{fig:spec} G and H) is won by
the latter one, leading to redder ($B-V$) and ($V-K$) colors for shallower
IMF slope.  Comparison of the ($B-V$) color drifts of the old and young
populations inspires us to search for the critical age where ($B-V$) color
changes from drifting blue to drifting red. We find that the ($B-V$)
colors of age less than 6 Gyr drift blue, but the ($B-V$) colors of age
greater than 8 Gyr drift red. Therefore, between 6 Gyr and 8 Gyr, the main
contributor of the IMF slope effect for ($B-V$) colors switches from
phase b stars to phase c stars.  

\begin{figure*}
\centering
\includegraphics [width=0.8\textwidth]{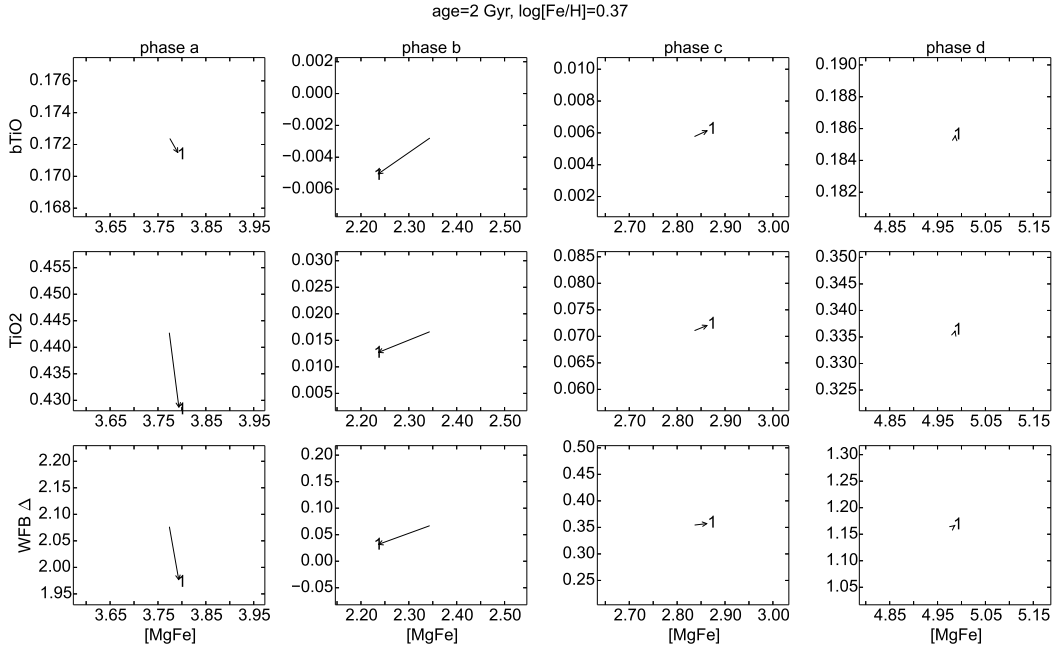} 
\caption[IMF slope drifts of each phase at young age]{IMF slope drifts
  of each phase at young age.}\label{fig:ind2}
\end{figure*}

Before ending this section, 
we calculate the indices of each phase (Figure \ref{fig:ind2}) to
explain the IMF slope effects. The drift magnitude ratios of phase
c over phase b stars has increased to about four times as the old
population. Given that phase c stars dominate the flux contribution in
most wavelength, bTiO, TiO2, WFB $\Delta$, and [MgFe] indices increase.


\section{ADF-IMF coupling}
\label{sect:coupling}
To explore the ADF-IMF coupling, we built a toy model in
which a hypothetical linear IMF$-$[M/H] relation is assumed.  This
linear relation is set by two points: $\alpha=2.35$ at
$\log {\rm [M/H]}=0$, and $\alpha=1.35$ at $\log {\rm [M/H]}=-1$.  Similar to
Paper I, we employ CSPs with single-burst ages but normal-width
composite abundance distribution functions (ADFs). Our normal-width
ADF matches well the average solar neighborhood ADFs and Milky
Way bulge ADFs in Paper I. To complete the models, the ADF-weighted Single Stellar
Populations (SSPs) are combined to construct CSPs, in which the IMF slope of each
mass bin matches the assumed IMF$-$[M/H] relation.  To
isolate the consequences of assuming a IMF$-$[M/H] relation, we also
construct another CSP models with the normal-width ADF, but instead of
assuming an IMF$-$[M/H] relation, we set the IMF to be always Salpeter
($\alpha=2.35$). We name the latter models  Constant-IMF CSPs (CCSPs),
and call the former models  Variable-IMF CSPs (VCSPs).  To verify
the robustness of our models, we assemble observed elliptical galaxy spectra
from  \citet{Graves2007}, \citet{Trager2008},
and \citet{Serven2010pdt}. Readers are referred to Paper I for sample
descriptions. Note that all the model and observed indices are
corrected to 300 km s$^{-1}$ resolution.

\begin{figure}
\includegraphics [width=0.5\textwidth]{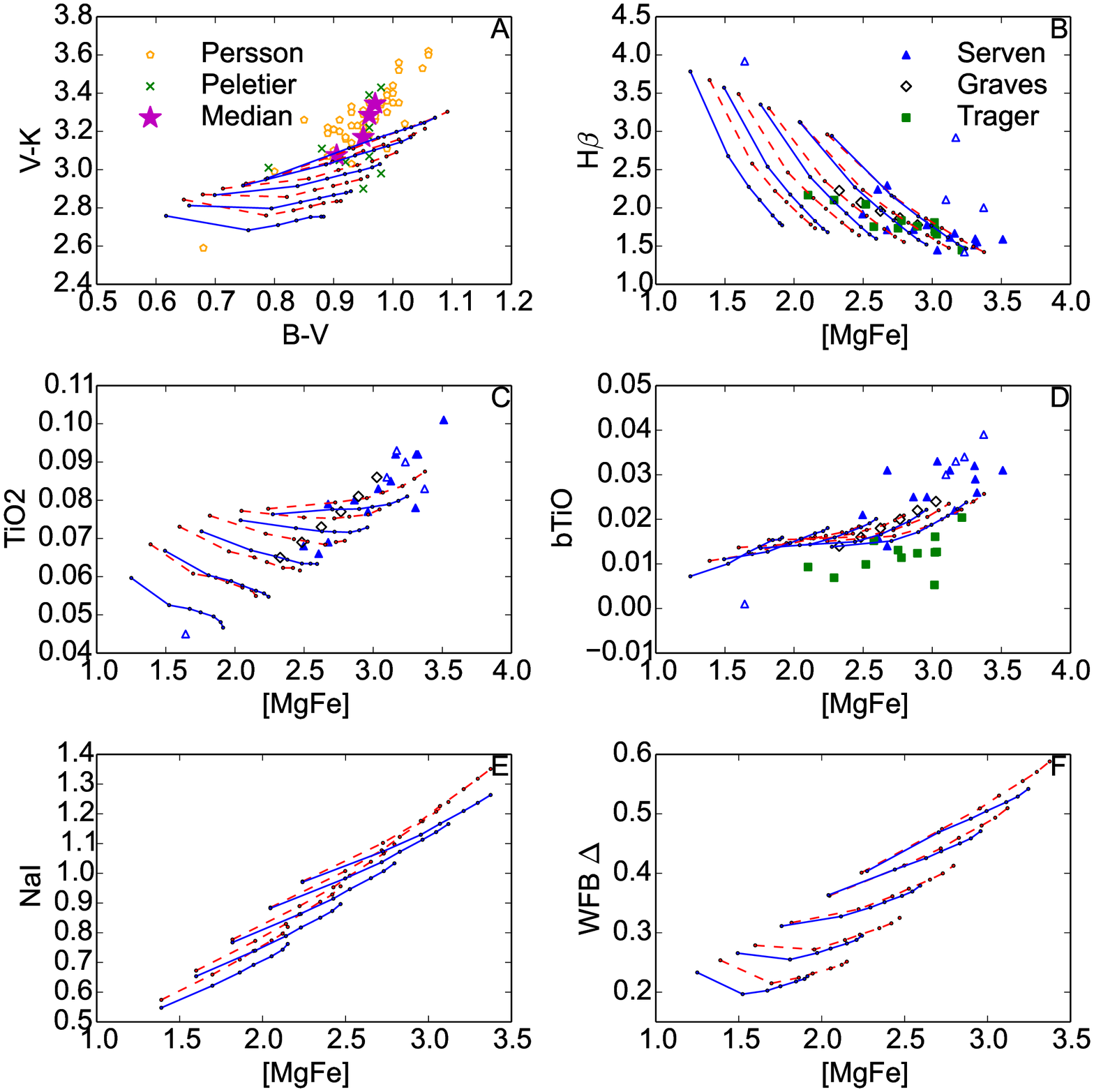} 
\caption[{CSPs with IMF$-$[M/H] dependence}]{The VCSPs (dashed
  lines) and CCSPs (solid lines) are described in
  \ref{sect:coupling}. In the top left panel, measurements from
  \citet{Persson1979} (open pentagons), and \citet{Peletier1989}
  (crosses) are shown. The medians of different velocity
  dispersions ($<$150, 150$-$200, 200$-$250, $>$250 km s$^{-1}$) are
  indicated as solid stars. In the optical index plots, Serven
  (triangles), Graves (open diamonds), and Trager (squares) 
  observations are shown. Five Balmer emission galaxies in the Serven
  sample (open triangles) have had additional corrections
  applied.}\label{fig:csp}
\end{figure}

First, we take a look at the optical-NIR color-color plots (Figure
\ref{fig:csp}A). The CCSPs and VCSPs are similar, but show
differences, especially for the
metal-poor populations. Since the photometric observations are not included in
the above samples, we retrieve the ($B-V$) and ($V-K$) colors from
\citet{Peletier1989} and \citet{Persson1979}. 
\begin{enumerate}
\item{\textbf{\citet{Peletier1989}:}}
In order to avoid focusing on only the cores of elliptical
galaxies, we choose the colors at maximum isophotal radius. These
colors are corrected for Galactic extinction before plotting.
\item{\textbf{\citet{Persson1979}:}}
We choose the well-tabulated ($B-V$) and ($V-K$) colors of field
galaxies. These colors have been corrected for Galactic extinction in
that paper.
\end{enumerate}
In Fig. \ref{fig:csp}A the two photometric samples are divided into
four bins in velocity dispersion ($<$150, 150$-$200, 200$-$250, $>$250
km s$^{-1}$) and we show the median colors of each bin as solid stars.
These medians confirm that massive elliptical galaxies tend have
redder ($V-K$) and ($B-V$) colors.  Note that the ($V-K$)
colors are only partially covered by the model grids. This
model-observation mismatch might be partly because CSPs with peak
[M/H]$>0.4$ is required, but we note that models with different
isochrones \citep{Worthey1994a} have no trouble reaching red enough
($V-K$) colors. This is due to a slightly stronger coupling between
metallicity and the temperature of the first-ascent giant branches in
the older models. The ($V-K$) miss is almost certainly a model
defect. We note that that flaw has little bearing on the results of
this paper which are differential in character.

Figure \ref{fig:csp}B is the H$\beta-$[MgFe] plot, which is 
an age-metallicity diagnostic diagram. We apply nebular emission
corrections for H$\beta$ to the Graves sample and Serven sample
following the 
recipe of \citet{Serven2010}. Five emission-corrected galaxies in the Serven
sample are labelled as open triangles, since larger uncertainties
might be expected for these galaxies in plots involving Balmer
features.  Generally speaking, the
observational data points from the three samples locate at the old,
metal-rich region, as what we expected.

Next, we pick several IMF-sensitive indices to
study the model variations. Though the Serven sample and the Trager
sample show moderate scatter compared with the model grids, it is encouraging to find the
observed data from \citet{Graves2007} are well-covered by our models,
and the variation trend looks reasonable.

In all the panels of Figure \ref{fig:csp}, we find that the VCSPs
appear more metal-rich than the CCSPs.  According to our IMF$-$[M/H]
relation, the IMFs of the metal-poor populations in the VCSPs are
shallower. We show that the shallower IMF model appear
dimmer than the standard IMF model in Figure \ref{fig:vkm}. Therefore,
the metal-poor populations appear dimmer in the VCSPs. Note that
metal-poor populations outshine their metal-rich
counterparts\footnote{If both populations have similar number of
  stars}, and thus the smaller integrated-light contribution from the
metal-poor populations leads to more
metal-rich appearing VCSPs.

\section{Discussion}

\subsection{Swapping Models}
To test the robustness of the model drifts, we also perform similar
experiments with the Flexible Stellar Population Synthesis (FSPS,
\citealt{Conroy2010}).  We arranged to utilize the FSPS with the same
age, IMF slope, and LMCO as our models. But the maximum metallicity of
the Padova$+$MILES option is $\log$[Fe/H]$=0.20$, which is different than
the metallicity we chose above (0.37). All the IMF-related
spectral changes might be age and metallicity sensitive.  Since the wavelength
coverage of MILES is 3600$-$7400 \AA, the \nai~and WFB $\Delta$ are
not available for analysis. To evaluate the AGB effect in FSPS, we
choose the original Padova TP-AGB treatment \citep{Marigo2007} as the standard model, and
compare it with the TP-AGB treatment of \citet{Conroy2010}. Note that
the latter TP-AGB treatment effectively reduces the NIR related colors
at age around 1 Gyr, and well fits the colors of star clusters in the
Magellanic Cloud (See their Figure 3). But both TP-AGB treatments have
similar descriptions at old ages, and thus we expect small AGB drift
for the old population in FSPS. This is different than the way we
change the AGB strength in $\S$\ref{sect:old}: the AGB strength is
always reduced to 80\% of the full strength, regardless of the age.

\begin{figure}
\includegraphics [width=0.5\textwidth]{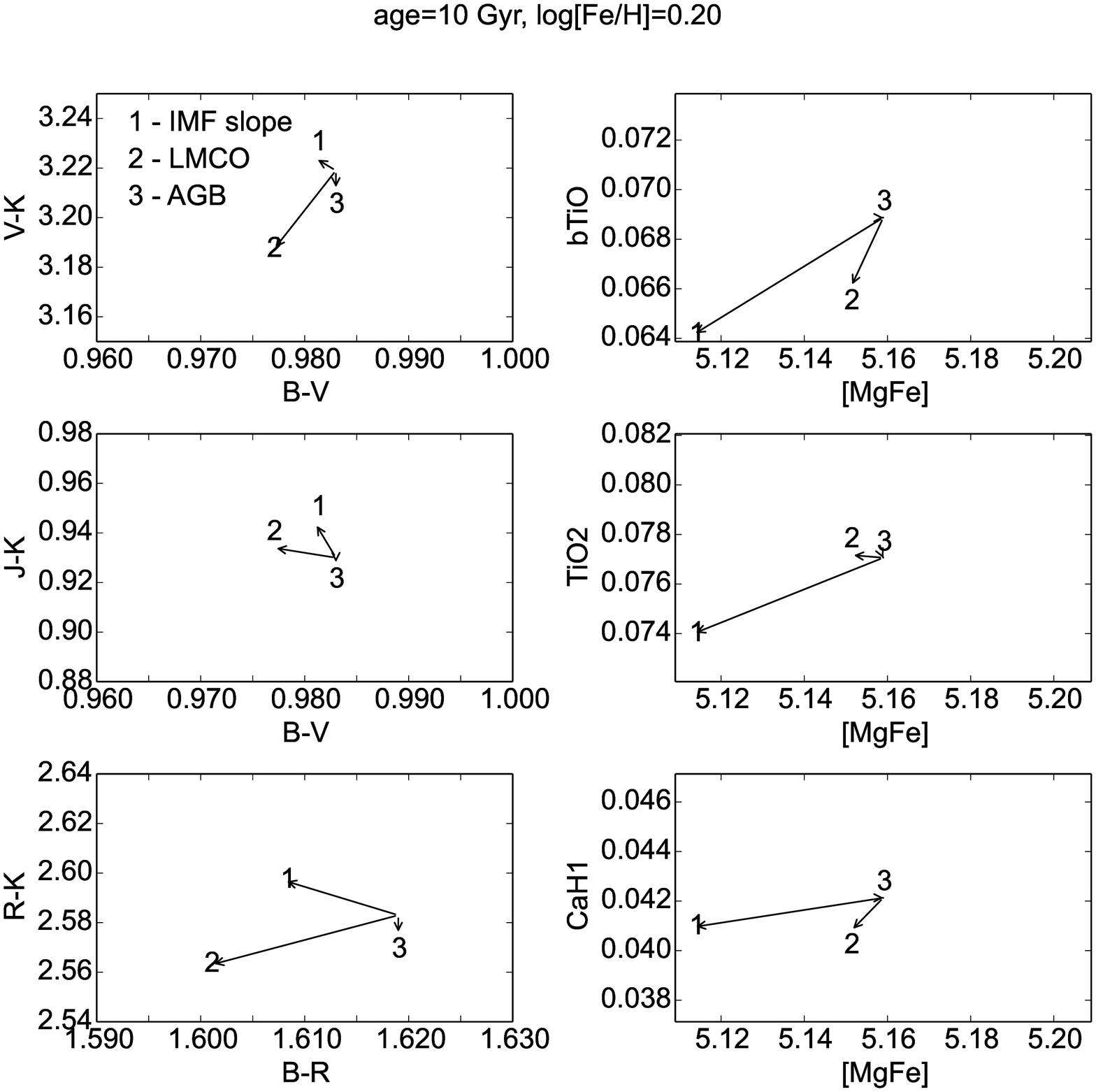} 
\caption[Color-color and index-index plots at old age using FSPS models]{Color-color and index-index plots at old age using FSPS models.}\label{fig:fd1}
\end{figure}

\begin{figure}
\includegraphics [width=0.5\textwidth]{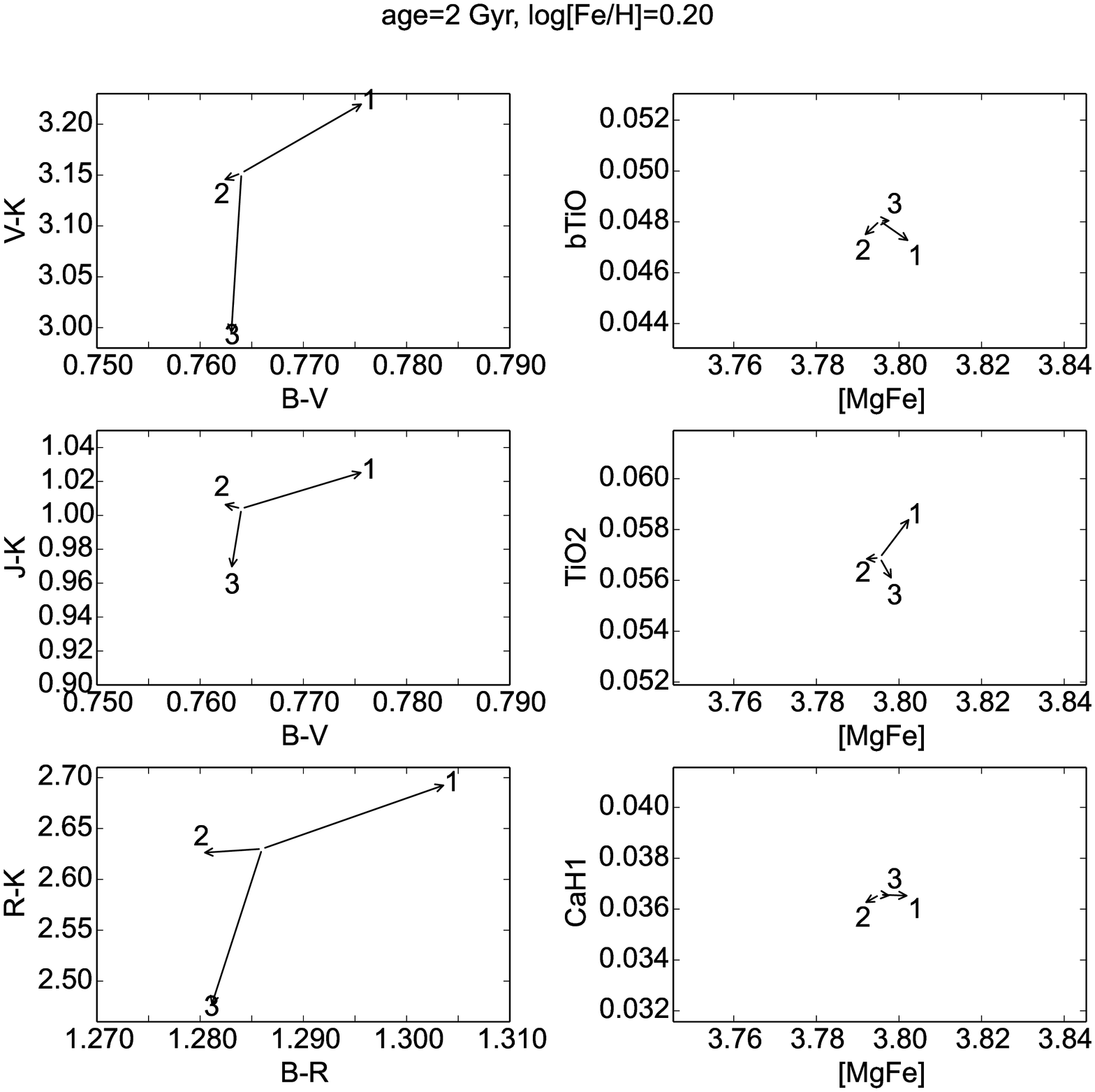} 
\caption[Color-color and index-index plots at young age using FSPS models]{Color-color and index-index plots at young age using FSPS models.}\label{fig:fd2}
\end{figure}

We show the color-color plots and optical index-index plots in Figure
\ref{fig:fd1} and \ref{fig:fd2}. To compare with the results from our
models, all the dynamic ranges are set to be the same as Figure
\ref{fig:qs} and \ref{fig:y}.  It is encouraging to find a lot of
similarities in the directions and magnitudes of the drifts between
two sets of models, but discrepancies still exist. For the old
population, all the AGB drifts are very small, as expected. The IMF
slope drifts seem smaller in the ($V-$K) colors, but show much bigger
magnitudes in the negative [MgFe] direction. Note that the LMCO drift
is no longer greater than the IMF slope drift for the TiO2 index: the IMF
slope effect is the leading effect, now.  Turning to the young
population model, the color drifts of LMCO and IMF slope seem robust
between two sets of models, but the AGB drifts are greater in the FSPS
models, due to the strong reduction of TP-AGB stars around 1
Gyr. Surprisingly, the index drifts are much smaller than our
models. This is possibly because of the different index
derivation routines of these two model sets, but we are not sure of that.

\subsection{Feasibility of Breaking the Degeneracy}
Exploring whether the apparent steep
IMF in massive elliptical galaxies might be partially due instead to the
decreasing number of AGB stars requires one to be able to find an
observational signature distinguish the two effects. We add to that
list the LMCO, of course, but also note that various age and
metallicity effects need to also be addressed, historically
derived from the
H$\beta-<$Fe$>$\footnote{$<$Fe$>$=(Fe5270 + Fe5335)/2 \citep{Gonzalez1993}.}
 or H$\gamma_{F}-<$Fe$>$ plot
 \citep{Trager2000a,Trager2000b,Tang2009}.

For a young elliptical galaxy, the color-color plots of Figure \ref{fig:y}
show substantial drifts ($\sim$0.03 mag) for the IMF slope and AGB effects. 
Note that the AGB and IMF slope effects vector oppositely in most of
the Figure \ref{fig:y} plots . Since our results should be interpreted
in a partial derivative sense, the opposite vectors in fact do nothing
to disentangle the IMF slope and AGB effects.  That is to say, the
degeneracy is still firm and hard to break in this case.

But for an old, metal-rich galaxy, the color-color plots of Figure
\ref{fig:qs} show the LMCO, IMF slope, and AGB effects are
distinguishable if the photometric accuracy 
is better than 0.02 mag.
The index-index plots reveal robust IMF slope drift in the [MgFe]
direction, reaching an amplitude of 0.02~\AA. This separates the
IMF slope effect from the LMCO effect, for the latter one causes much
smaller drifts in the [MgFe] direction. Note that the [MgFe] index is
insensitive to [$\alpha$/Fe], giving the [MgFe]-related plots advantage
to different alpha-enhanced environment.

We find therefore that it is practical to lift the degeneracies using
optical-red spectra and optical-NIR photometry if the
measurements are accurate enough. We estimate these accuracies should
be achieved: (1) dust attenuation $\sigma_{A_{V}}<$0.01 mag;
(2) photometric accuracy $\sigma<$0.01 mag; (3) index accuracy
$\sigma<$0.02 \AA. As stellar photometric accuracy reaches
milli-mag these days \citep{Clem2007}, an accuracy of 0.01 mag is feasible for
nearby galaxies if one is precise about the observational complications. 
To list a few of them: dust attenuation, both Galactic and in-situ; sky subtraction for extended
sources; aperture matching  for spectroscopic and photometric extractions.

We also notice that the \nai~index seems to be insensitive to the AGB
effect for both the old and the young populations, in spite of \nai's red
central wavelength. A look at the index fitting functions at the
stellar level reveals this insensitivity
may relate to the similar \nai~indices of giants and dwarfs. The \nai~
index might be a way out of the ADF-AGB-IMF masquerading.

\begin{figure}
\includegraphics [width=0.5\textwidth]{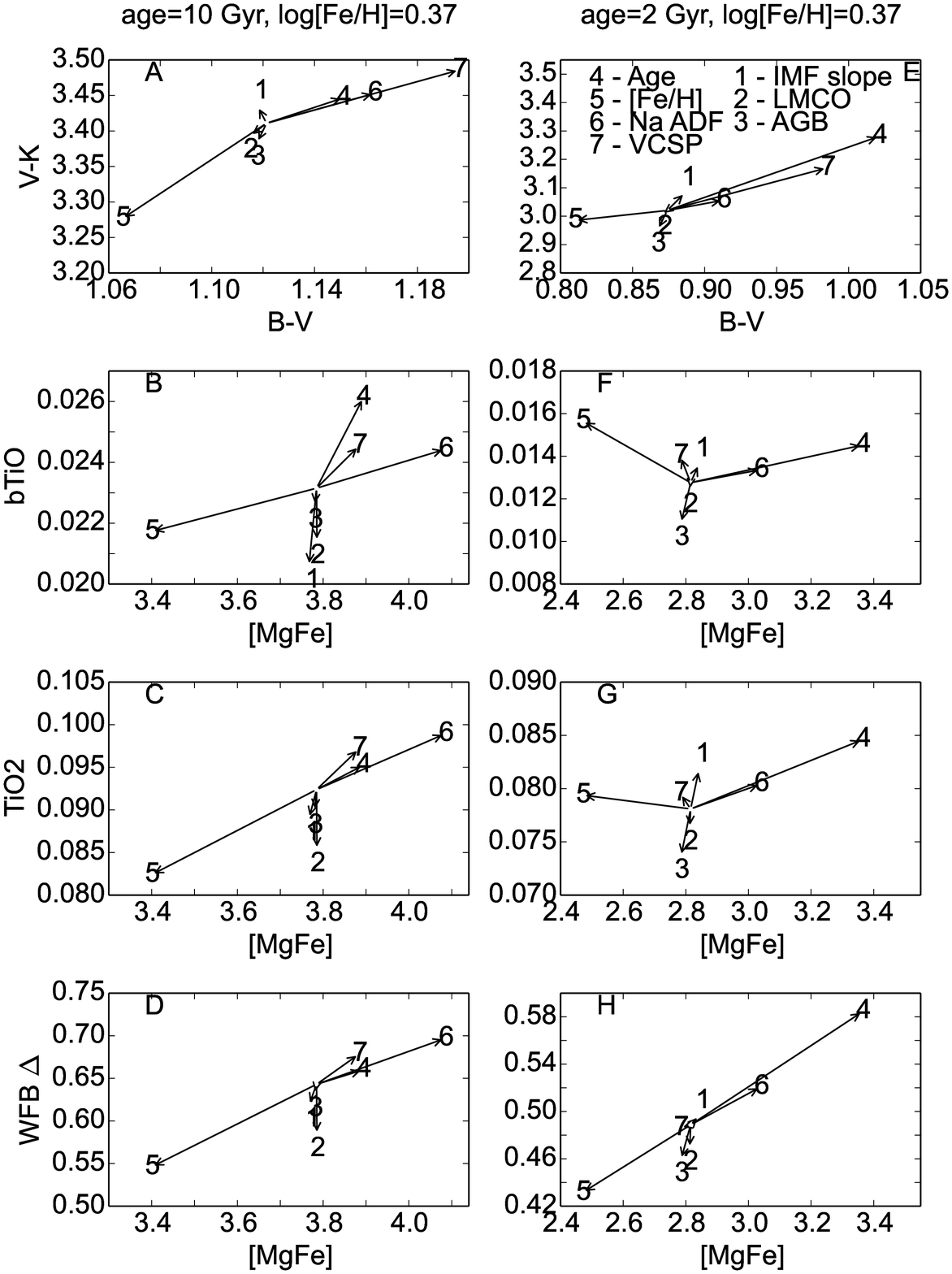} 
\caption[Symposium of multiple effects]{To visualize effects that may
  impact the SSP parameter determination, IMF slope, LMCO, AGB contribution, age,
  metallicity, ADF, and VCSP effects are labelled as vector 1 to 7, respectively.}\label{fig:all}
\end{figure}

\subsection{Symposium of Multiple Effects}
\label{sect:all}
Age and metallicity effects are generally strong, but also degenerate
in most of the colors and indices \citep{Worthey1994a}.  To obtain a
bigger picture besides the three effects that we discuss in this
paper, we summarize the effects concerning the IMF slope, LMCO, AGB contribution, age,
metallicity, ADF, and VCSP in Figure \ref{fig:all}. They are labelled
as arrow 1 to 7, respectively. The parametrization of the first three
effects is described in $\S$\ref{sect:mo}. The age effect is estimated
by calculating SP models of age$=$12 Gyr, $\log$[Fe/H]$=$0.37, and
age$=$4 Gyr, $\log$[Fe/H]$=$0.37 ($\Delta$ age=$+2$ Gyr), while the
metallicity effect is estimated by calculating SP models of age$=$10
Gyr, $\log$[Fe/H]$=$0.185, and age$=$2 Gyr, $\log$[Fe/H]$=$0.185
($\Delta \log$[Fe/H]$=-0.185$).  To investigate the effects concerning
CSPs, we move the models with normal-width ADF, peak [M/H]$=0.4$, and
constant IMF slope to the center. The ADF effect is shown as a displacement from the
normal-width ADF CSPs to the narrow-width ADF CSPs, and the ADF-IMF coupling is
represented by a vector from the CCSPs to the VCSPs.

Age and metallicity effects are parallel and degenerate as
expected. They are joined by two new effects concerning the CSPs: the
ADF and VCSP effects. In fact, the ADF effect is the same as the ``red
lean'' effect (Paper I) where narrower ADF appears more metal-rich
than a wider one. The reasons for the VCSP effect are
explained in $\S$\ref{sect:coupling}. Careful readers may find that
the ADF and VCSP effects both originate from the suppression of
metal-poor populations, which leads to a metal-richer appearing
integrated-light model.  The similar origins and vector directions
drive us to segregate the age, metallicity, ADF and VCSP effects into one
group, and label them as group I effects.  
Group II effects are IMF slope, LMCO, and AGB contribution. Group II effects vector
almost orthogonally to group I effects.  Though group I effects
are more prominent in magnitude, the orthogonality suggests
that observations to isolate group II effects are still feasible.

\subsection{Recovering [X/Fe], IMF slope, LMCO, and AGB Percentage}
\label{sect:recover}
\citet{Worthey2014} showed our efforts to recover elemental abundances
from observed spectra. In this work, we upgrade our inversion program by
replacing the SSP models with normal-width ADF CSP
models and adding the IMF slope, LMCO, and AGB percentage (AGB\%)
parameter determination. 

\begin{figure*}
\includegraphics [width=0.75\textwidth]{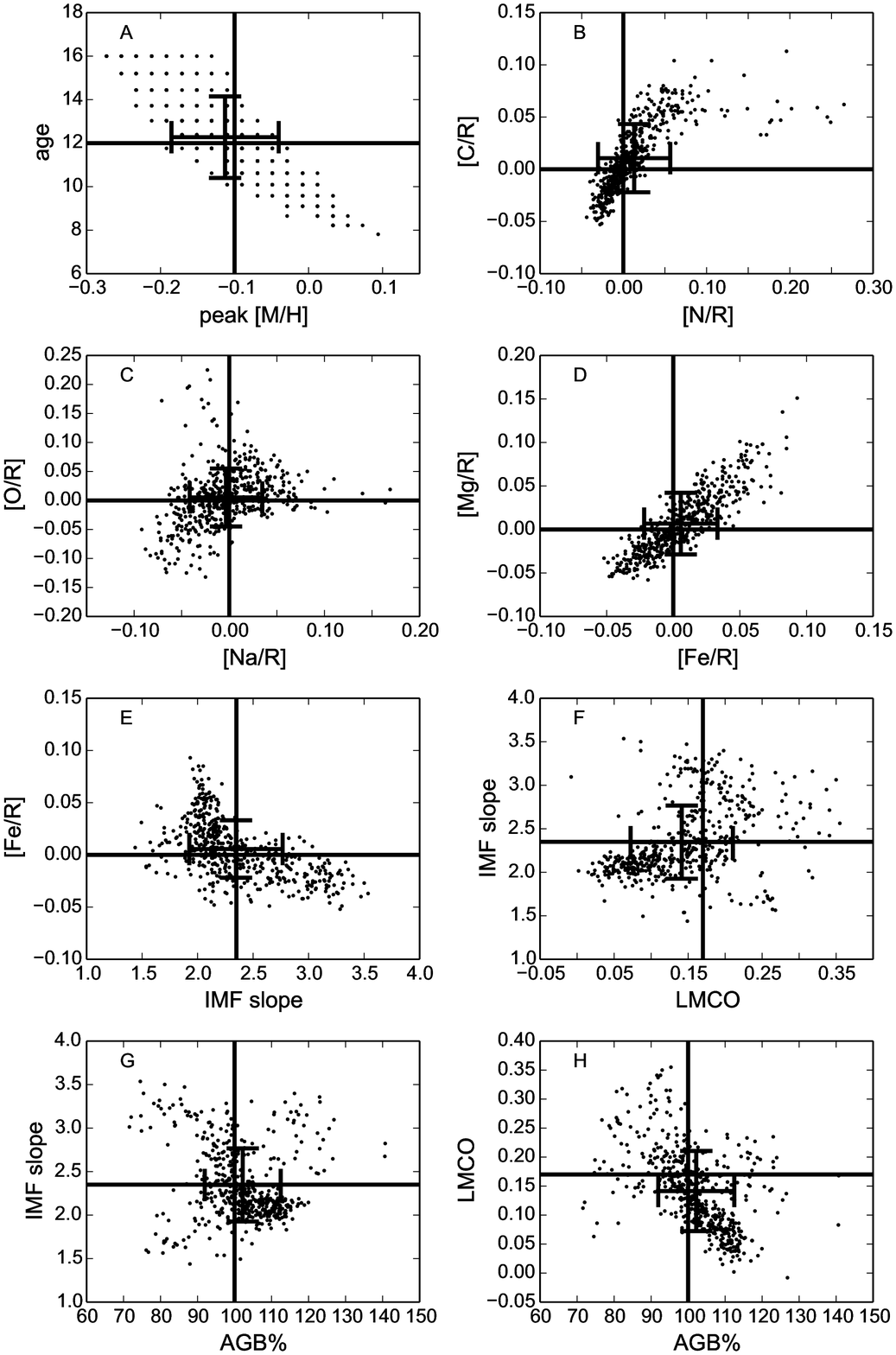} 
\caption[Recovering parameters]{The recovered values (scattered dots) are
  compared with the input values (solid lines). The mean recovered
  values and their corresponding standard deviations are labelled with
  error bars.}\label{fig:mc}
\end{figure*}

To examine the uncertainties of our inversion program, 500 mock galaxy
spectra are constructed by Monte Carlo simulation. First, a model is
produced with known age, abundance, IMF slope, LMCO, and AGB
fraction. In this model the indices and colors are altered by random
numbers which have a normal distribution.  The standard deviation of
the normal distribution function is set to 0.7$\sigma_{std}$, where
$\sigma_{std}$ is a vector composed for reasonable guesses for the
observational uncertainty in each color or index (Table
\ref{tab2}). We supply the inversion program with each of these mock
galaxy spectra in turn, and compare the recovered values with input
values.  Figure \ref{fig:mc} show the recovered and input values of
models at age = 12 Gyr, peak [M/H] = $-0.1$.

\begin{table*}
\begin{minipage}{\textwidth}   
\caption{\label{tab2} Baseline Assumed Observational Uncertainty in Each Color or Index ($\sigma_{std}$)}
\begin{center}
\begin{tabular}{cccccccccccccccc}
\hline \hline
\ & Index & $\sigma$ & Unit & \  & Index & $\sigma$ & Unit & \  &
Index & $\sigma$ & Unit & \  & Index & $\sigma$ & Unit \\
\hline 
1&CN1       &   0.005&mag &24 &H$\delta_F$   &   0.121 &\AA& 47&Cr3594     &   0.131 &\AA&70&bTiO$_{STKC}$\footnote{From \citet{Spiniello2014a}}  &   0.005 &mag\\
2&CN2       &   0.006 &mag& 25&H$\gamma_F$   &   0.114 &\AA& 48&Cr4264     &   0.157 &\AA& 71&aTiO$_{STKC}$  &   0.003 &mag\\
3&Ca4227     &   0.081 &\AA& 26&H$\alpha$    &   0.114 &\AA& 49&Cr5206     &   0.077 &\AA& 72&CaH1$_{STKC}$  &   0.004 &mag\\
4&G4300      &   0.155 &\AA& 27&CO4685     &   0.339 &\AA& 50&Mn3794     &   0.230 &\AA& 73&CaH2$_{STKC}$  &   0.003 &mag\\
5&Fe4383     &   0.225 &\AA& 28&CO5161     &   0.083 &\AA& 51&Mn4018     &   0.325 &\AA& 74&\nai$_{STKC}$   &   0.243 &\AA\\
6&Ca4455     &   0.128 &\AA& 29&CNO3862    &   0.221 &\AA& 52&Mn4061
&   0.147 &\AA& 75&TiO2$_{SDSS}$\footnote{From \citet{LaBarbera2013}} &   0.003&mag \\
7&Fe4531     &   0.190 &\AA& 30&CNO4175    &   0.285 &\AA& 53&Mn4757
&   0.135 &\AA& 76&H$\beta$emiss\footnote{From \citet{Gonzalez1993}}    &   0.090 &\AA\\
8&C$_2$4668    &   0.319 &\AA& 31&Na8190     &   0.140 &\AA& 54&Fe4058     &   0.173 &\AA&77&\oiii emiss1   &   0.139 &\AA\\
9&H$\beta$     &   0.131 &\AA& 32&Mg3835     &   0.156 &\AA& 55&Fe4930     &   0.164 &\AA&78&\oiii emiss2   &   0.091 &\AA\\
10&Fe5015     &   0.285 &\AA& 33&Mg4780     &   0.156 &\AA& 56&Co3701
&   0.135 &\AA& 79&H$\beta_0$\footnote{From \citet{Cervantes2009}}   &   0.159 &\AA\\
11&Mg1       &   0.003 &mag& 34&Si4101     &   0.096 &\AA& 57&Co3840     &   0.142 &\AA& 80&WFB $\Delta$\footnote{From \citet{Whitford1977}}  &   0.132 &\AA\\
12&Mg2       &   0.004 &mag& 35&Si4513     &   0.164 &\AA& 58&Co3876     &   0.216 &\AA& 81&($NUV-U$)      &   0.020 &mag\\
13&Mgb       &   0.136 &\AA& 36&Cahk       &   0.350 &\AA& 59&Co7815     &   0.206 &\AA& 82&($U-V$)        &   0.010 &mag\\
14&Fe5270     &   0.151 &\AA& 37&Ca8542     &   0.125 &\AA& 60&Co8185     &   0.146 &\AA& 83&($B-V$)        &   0.005 &mag\\
15&Fe5335     &   0.172 &\AA& 38&Ca8662     &   0.106 &\AA& 61&Ni3667     &   0.215 &\AA& 84&($V-I$)        &   0.005 &mag\\
16&Fe5406     &   0.130 &\AA& 39&Sc4312     &   0.115 &\AA& 62&Ni3780     &   0.261 &\AA& 85&($I-J$)       &   0.010 &mag\\
17&Fe5709     &   0.106 &\AA& 40&Sc6292     &   0.167 &\AA& 63&Ni4292     &   0.131 &\AA& 86&($I-K$)        &   0.010 &mag\\
18&Fe5782     &   0.102 &\AA& 41&Ti4296     &   0.112 &\AA& 64&Ni4910     &   0.128 &\AA& 87&($V-K$)        &   0.015 &mag\\
19&NaD    &   0.127 &\AA& 42&Ti4533     &   0.087 &\AA& 65&Ni4976     &   0.116 &\AA& 88&($I-L$)        &   0.020 &mag\\
20&TiO1      &   0.003 &mag& 43&Ti5000     &   0.196 &\AA& 66&Ni5592     &   0.117 &\AA&89&IRAC ($3.6-8.0$)    &   0.020 &mag\\
21&TiO2      &   0.003 &mag& 44&V4112      &   0.282 &\AA& 67&Ba4552     &   0.081 &\AA& 90&WFC3 ($225w-555w$)  &   0.020 &mag\\
22&H$\delta_A$   &   0.185 &\AA& 45&V4928      &   0.097 &\AA& 68&Ba4933     &   0.113 &\AA& 91&($U-K$)        &   0.020 &mag\\
23&H$\gamma_A$   &   0.195 &\AA& 46&V6604      &   0.095 &\AA& 69&Ba6142     &   0.100 &\AA&92&($NUV-K$)      &   0.020 &mag\\
\hline
\end{tabular}
\end{center}
\end{minipage}
\end{table*}

In Figure \ref{fig:mc}A, the age and peak [M/H] are 
quantized in the current inversion algorithm, since these two basic parameters are determined by
closest-match rather than a smooth interpolation within the model
grid. Age-metallicity degeneracy is the reason for the
anti-correlation found in the first panel. In Figure \ref{fig:mc}B, we
see tight correlation between [C/R] and [N/R], since these two
elements are largely determined by common indices, the CN bands around 4100
\AA. In Figure \ref{fig:mc}F, \ref{fig:mc}G, and \ref{fig:mc}H, the
correlations among IMF slope, LMCO, and AGB\% are generally
loose. This implies the degeneracies among these three parameters are,
for the most part, broken by our selection of indices.  Though modest
scatter exist in the recovered values, all of the mean recovered
values agree with the input values inside the error range without
alarming systematics. Some of the mean recovered values are especially
close to the input values, e.g., IMF slope, [O/R] and
[Na/R]. This success encourages us to apply our inversion program to
observed galaxy spectra and colors in the future.

\section{Summary}
\label{sect:con}

A modeling study of various new parametric effects on integrated
light, begun in Paper I by exploring the effects of the Abundance
Distribution Function (ADF), continues on in this paper to explore two
underexplored effects that might mimic IMF slope variation: the LMCO
and the AGB strength. If we hypothesize that the apparently-steep IMF
in massive elliptical galaxies might be partially due to the
decreasing number of AGB stars, the question is can we ever know it,
given the stellar population degeneracies. The magnitudes of the IMF
slope, LMCO, and AGB effects are smaller than age and metallicity
effects, but these two groups of effects vector almost orthogonally in
diagnostic diagrams, which bodes well for measuring the
\textit{combination} of slope/LMCO/AGB effects. Internal to that trio,
we explore the degeneracies and find that it is very difficult to
distinguish steepening IMF from decreasing AGB strength for young,
metal-rich populations. However, the slope/LMCO/AGB degeneracies can
be lifted for old (age$\approx$10 Gyr), metal-rich populations using a
combination of optical-to-near infrared photometry and spectroscopy.
The disentanglement happens, however, at an observationally
challenging level ($\approx$0.02 mag).

We fragmented our models into different evolutionary phases to isolate
and discuss the leading factors for the IMF-related effects.  We also
investigated a series of models with an ADF-IMF coupling in which
metal-rich populations favor low mass star formation. Models with
ADF-IMF coupling appear more metal-rich than the noncoupled models,
strongly resembling a modest narrowing of the ADF width, which has the
same ``red lean'' effect. Attempting to recover the magnitude of an
ADF-IMF coupling is very challenging from integrated light alone.

We upgraded our inversion program \citep{Worthey2014} by replacing the
SSP models with normal-width ADF CSP models and adding the IMF slope,
LMCO, and AGB percentage parameter determination. We estimated
uncertainties on parameter estimation from Monte Carlo simulations,
and find no significant systematic drifts, though of course we see
many partial parameter degeneracies. The degeneracies among the
IMF-related parameters can be broken by our selection of indices for
old stellar populations. This success encourages us to seek
appropriate observational material to apply our techniques to observed
galaxy spectra and colors in the future.

\section{Acknowledgements} 
The authors would like to thank I. Chilingarian for access to his
improved version of the CFL library.

\bibliographystyle{mn}
\bibliography{imfqs}

\label{lastpage}

\end{document}